# CVR: A Continuously Variable Rate LDPC Decoder Using Parity Check Extension for Minimum Latency

Sina Pourjabar, Gwan S. Choi
Department of Electrical and Computer Engineering Texas A&M University, College Station, TX
Email: {pourjabar, gwanchoi}@tamu.edu

*Abstract*— This brief presents a novel IEEE 802.16e (WiMAX) based decoder that performs close to the 5G code but without the expensive hardware re-development cost. The design uses an extension of the existing WiMAX parity check code to reduce the processing latency and power consumption while keeping the decoder throughput at maximum. It achieves similar Frame Error Rate (FER) compared to 5G (0.1dB off), and most notably the error curves trend down like 5G instead flooring. At FER= $10^{-3}$ there is 0.1 dB gain in the FER code performance compared to WiMAX. An implementation of the design is a modified version of the existing fully-parallel WiMAX decoder that supports multi-rate codeword size and reduces latency by 33%. Additionally, for SNR greater than 3dB, decoding only the shorter code reduces the power consumption by 36%.

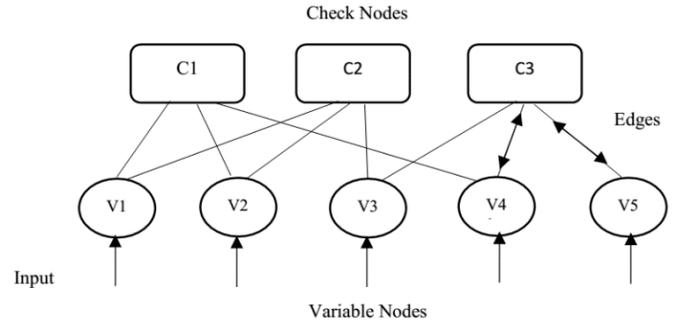

Fig.1. LDPC Tanner graph

## I. INTRODUCTION

LOW Density Parity Check (LDPC) codes were first introduced by Gallager in 1960 [1]. One major advantage of these codes is their fully parallel architecture that gives a significant boost to the throughput in theory. However, due to the complex structure of these codes, they were difficult to implement until the rediscovery of these codes in 1996 by MacKay[2]. With further increase of data rate in newer telecommunication protocols such as 5G NR, LDPC codes are playing the key role in reaching the Shannon channel capacity while allowing more flexibility on power consumption. LDPC code is defined with a parity check matrix H that is mapped to a bipartite graph. As figure 1 shows, each graph is made of Check Node Units (CNU) and Variable Node Units (VNU). Decoding of LDPC codes is based on a voting system in which the correct value of an information bit is decided by other bits that share the same CNU. Decoding algorithm can be performed in two different ways; hard decoding or soft decoding. In hard decoding, input to every VNU is a one-bit value that is used for deciding whether to flip the value of another bit in a common check node or not. The decoding process is finished whenever the decoder converges to a final value by satisfying the $CH^T = 0$, (C is the received message) or the decoder reaches the maximum iterations limit defined by the user. The output of the $CH^T$ equation is also known as syndrome. Syndrome value is an indicator of total number of uncorrected errors in a received message. The hardware implementation for a hard decoder is simpler than soft decoder and is mostly used in designs where the frame size is small. On the other hand, for a decoding process which deals with larger frame size with low SNR conditions, it will take more iterations for a hard-decoding algorithm to converge. In these circumstances, soft decoders are utilized in which primary input to variable nodes are made of Log Likelihood Ratios (LLR) or probabilities. In BPSK modulation, an LLR with a positive value is mapped to logical zero and the negative value of LLR is mapped to a logical one. After each iteration, extrinsic probabilities are getting updated. The certainty of an information bit value is amplified as the absolute value of each LLR increases until the final decoded message is obtained by satisfying the $CH^T = 0$ condition. In comparison to hard decoding, with a cost of a more complex hardware, the number of iterations can be reduced significantly. This method is also known as belief propagation [1] in which LLRs are calculated through the Sum-Product (SP) decoder. In SP decoding theupdate equations for VNs and CNs are respectively as follows:

$$M_{ij} = LLR(j) + \sum_{k=1, k \neq i}^{d_j^v} E_{kj} \qquad (1)$$

$$E_{ij} = 2\tanh^{-1}(\prod_{k=1, k \neq j}^{d_i^c} \tanh(\frac{E_{ik}}{2})) \qquad (2)$$

in which $M_{ij}$ is the message sent from VN in column j to the CN in row i, and similarly $E_{ij}$ is the updated message sent from CN to VN. However, since the architecture of SP decoder is complex another algorithm called Min-Sum decoding is introduced [3]. Min-Sum decoding reduces the hardware complexity in comparison to SP, but since there is an overestimation of updated values in CNU, the total number of iterations would increase. Thus, an offset value for canceling the overestimation in CNU is added to the design. This is also known as Offset Min-Sum decoder (OMS) [4]. In OMS, updating the check node equation can be approximated to:

$$E_{i,j} = \prod_{k=1, k \neq j}^{d_i^c} sign(M_{ik}).max(min|M_{ik}| - \beta, 0) \quad (3)$$

The implantation of OMS can be done by implementing, fully parallel [5], fully serial [6], partially parallel [7] or layered decoding [8]. The focus of this brief is in fully parallel architectures. Fully parallel design is appropriate from the processing speed perspective since it can converge using fewer iterations. Creating a multi-stage pipelined architecture in LDPC decoder contributes to even faster throughput with every clock cycle. However, because of the nature of pipelining in adding extra registers, the first frame of the decoder suffers an increased amount of delay in comparison to a non-pipelined design. One objective of this brief is looking for a technique to compensate for this delay.

Puncturing is a method initially introduced for Viterbi decoders and convolutional codes by [9]. In puncturing, the transmitter punctures part of the parity bits and sends rest of the code, and therefore a higher rate code emanates from a lower rate code [10, 11]. Puncturing is useful when working with high throughput transmission to make the most of the bandwidth without making the encoder and decoder architecture vastly complex. It also adds more flexibility to the hardware in terms of working with different code rates based on the channel conditions. However, as discussed in [12] puncturing will increase the gap with the channel capacity. This is an unwelcome phenomenon, and with a nosier channel condition it becomes worse. Another method to make a rate compatible decoder is by use of parity extension [12, 13]. In terms of parity check construction, extension follows an opposite way of puncturing. Extra parity bits are added to an already existed parity check matrix (PCM) or embedded mother code, thus creating a lower rate code. The new lower rate code shows better Frame Error Rate (FER) and Bit Error Rate (BER) in comparison to the mother code. These codes are used in type two Hybrid ARQ in which extra parity bits are sent by the transmitter only in case the decoder is unsuccessful in correcting the errors [14]. In the case of LDPC, depending on the structure of the extension in the PCM the code performance can be drastically enhanced [12, 15]. For the new 5G standard, multiple new sets of PCMs are introduced [16]. The largest set size of 5G has a codeword size of 25k in comparison to 2304 bits used in WiMAX. Theoretically, 5G is supporting throughputs up to 20Gbps [17]. For rate compatibility and having strong resilience in low SNR conditions, extension is also included in 5G parity matrix.

In this brief, a new approach for extending a parity check matrix based on the WiMAX ¾ rate parity is introduced. The extended parity design outperforms the standard half rate WiMAX with the same code length and the FER gets closer to the performance of the 5G parity code. Similarly, with the help of the new PCM and modifying the WiMAX hardware architecture, the new implementation benefits from a lower latency for any pipelined architecture. Consequently, it decreases the total amount of delay while reducing power consumption in high SNR conditions.

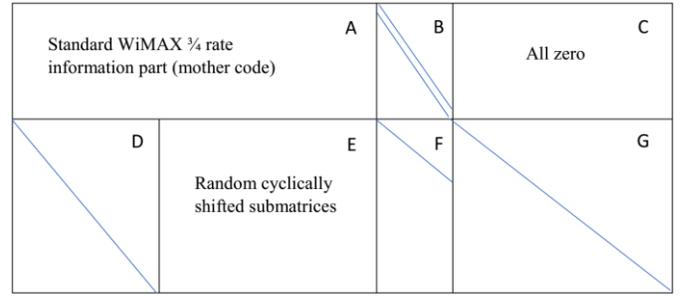

Fig. 2. Proposed parity extension structure

## II. DESIGN

Creating a high performing parity extension matrix is essential. Undoubtedly, increasing the number of parity bits contributes to improving the code performance. The main question is: to what extent does the code performance benefit from parity extension? To answer this question, a ¾ rate 144x576 WiMAX PCM is employed and extended to half-rate 432x864. It is shown that not only the chosen extension structure presents a better result than a 144x576 PCM but it also outperforms a non-extended half- rate WiMAX parity with the same size of 432x864. The extension method shown in Fig. 2 is applied on ¾ rate parity check WiMAX to extend it into a half-rate matrix.

Part A is the information part of the matrix Part B is the dual diagonal parity part that is already defined in the WiMAX standard. Hence, part A and B together are the standard WiMAX parity check matrix. Part C is an all-zero matrix. Part D is an identity matrix equal to the size of the number of check nodes in the standard WiMAX parity matrix. Part E is a randomly cyclically shifted matrix with this condition that each row consists of five to six 1's. Part F is another matrix that is placed exactly under part B and has the same number of columns. Each column and row in part F can have only one connection. Finally, the parity part of the extension matrix, part G, is an identity matrix. To be exact, simulations shows that part D should always be an identity matrix otherwise the code performance would reduce. Similarly, for part F, having only one connection in each row and column achieves the best possible performance. The only partition that is more flexible to modification is part E, and it was found that creating cyclically shifted submatrices results in lower error rates in decoding. Choosing G as an identity matrix not only improves the code performance in comparison to double diagonal parity extension, but also makes the encoding process easier while selecting among any arbitrary code rates. This parity matrix structure for the extension of a WiMAX code shows better performance in terms of having lower FER and BER. Fig. 3 does a comparison of three half-rate parity check matrices: 1. WiMAX with a codeword size of 864 2. Extended WiMAX with a codeword size 864 and 3. Half-rate 5G parity with a codeword size of 792. (5G does not support 864 bits half-rate codeword). A maximum of 21 iterations is performed for each frame. As presented, overall the 5G code is superior in overall FER and BER performance. The proposed extended technique reaches a higher BER in a lower SNR situation in comparison to a half-rate WiMAX. However, in higher SNR values (SNR> 2.6) the extended WiMAX BER graph crosses the WiMAX BER graph and demonstrates a lower BER. In the FER graph, at FER=$10^{-3}$ the extended version of WiMAX is performing better by 0.1 dB compared to the standard half-rate WiMAX. In addition, the difference between the FER in extension and 5G code is equal to 0.1 dB. The reason that extension illustrates worse BER but better FER in lower SNR situation is because most of the uncorrected errors in the extension method are accumulated in the first few frames while errors in standard WiMAX are dispersed throughout all frames. These graphs prove it is possible to create a new extended matrix based on an existing standard, that in identical conditions surpasses the standard WiMAX in terms of performance, and also performs closely to the newly introduced 5G standard.

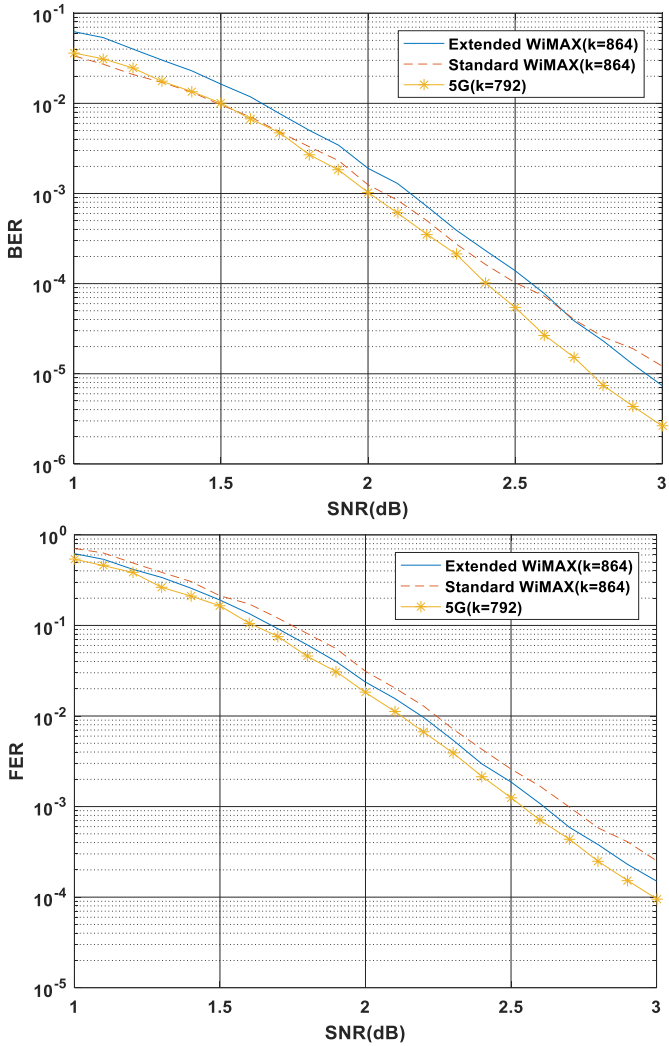

Fig. 3. Bit Error Rate and Frame Error Rate comparison of the proposed extension method with standard WiMAX and 5G.

III. CONTRIBUTIONS

The following section will discuss the improvements made

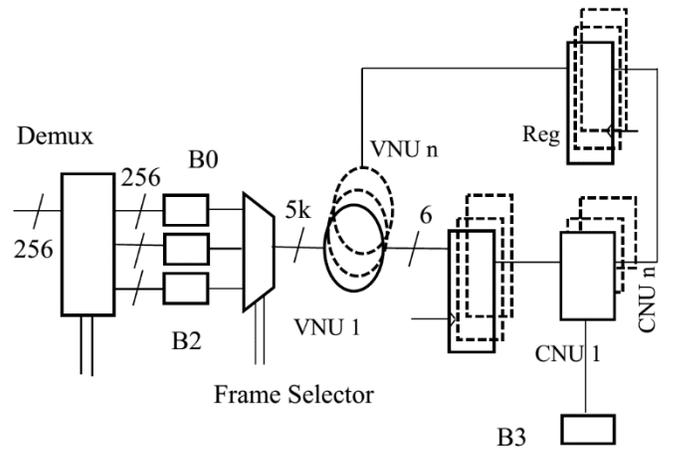

Fig. 4. Pipelined design for processing two frames simultaneously

in performance, processing delay and power consumption.

A. Delay Reduction

In a hardware implementation, one advantage of this design is that the decoder does not need to wait for the whole frame to be received. As soon as the small mother code fragment is acquired it can start decoding the first frame. In good SNR situations, there may be no need to receive the extra parity bits of the first frame. This way, the decoding on the first frame can start earlier, and in the meantime the transmitter can start sending the second frame. By the time the decoding of the first frame is completed, the second frame is ready to be decoded. Therefore, by shifting the whole transmission and decoding timeframe the total amount of latency is reduced.

B. Power Consumption

Generally, increasing a decoder's throughput escalates the total consumed power. This is one of the concerns related to flooding schedule used in fully parallel designs. To compensate for this power increase, different scheduling algorithms such as layer decoding have been proposed. Although they help save power, they reduce the decoding speed as well. The hardware implementation of the current extension parity matrix in good SNR situations could disable part of the CNUs and VNUs and save a significant amount of power without sacrificing the decoding speed. With the assumption of having a high SNR, a power estimation based on the FPGA implementation shows that the new decoder can decrease the power consumption up to 36% when it is set to decode the shorter code length of 576 compared to 864. The multi-rate compatibility of this design makes it suitable to be further optimized to work with any amount of parity checks when in need.

IV. ARCHITECTURE

As mentioned earlier, 5G is supporting a maximum throughput of 20 Gbps. The biggest codeword size is 25344 bits with the information part of 8448 bits. To be able to support this massive block size and processing speed, having an architecture with a high degree of parallelism is necessary. Therefore, for our design a multi-rate fully parallel architecture is chosen. It is also possible to have control over the power and processing speed. The proposed IO interface and the pipelined decoder are are shown in Fig. 4. Using a bus width of 256 bits, generated LLRs are transferred into 3 buffers B0, B1 and B2 with the help of a demux. As Fig. 5(a) exhibits, initially the buffer B0 gets filled in with the first frame data. To reduce the stall time of the decoder for the first frame, B0 only receives the shorter codeword and as soon as LLRs are received the decoder can start decoding. In the meantime, buffer B1 starts receiving the second frame with the longer codeword size for better decoding performance. When B1 is also ready, the decoder can decode two frames simultaneously at each clock cycle. Likewise, B2 is fed with the third frame so that it can be replaced once the data in one of the buffers is decoded successfully; as a result, the stall time for the next frame can be minimized. Once the decoding is finished the output is sent to the buffer B3. The decoder is quantized to 6 bits with the maximum number of iterations set to 21. The CNUs corresponding to the mother code with the higher rate are similar to a non-extended fully parallel decoder (Fig. 6(a)). The CNUs associated with the extra parity bit equations as shown in Fig. 6(b) are controlled with an enable signal that can be triggered with a simple frame counter or a code rate controller. the controller uses information from the previous decode cycles as well as SNR from the front-end receiver circuit. Accordingly, more control over the decoding speed as well as the consumed power for different applications is attained.

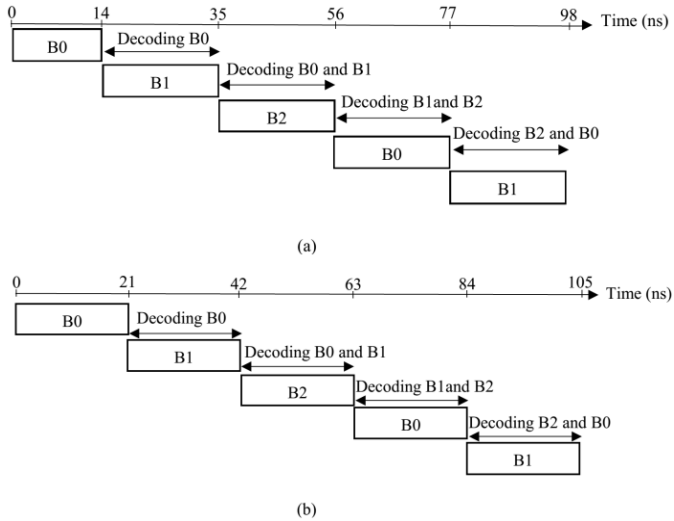

Fig. 5. Timing diagram of the pipelined design. (a) For the first frame, buffer B0 receives the shorter length code. (b) Conventional decoding scheme. The design can switch between (a) and (b).

Table I
FPGA implementation result

| Resources | Used | Utilization |
|---|---|---|
| Slice | 46669 | %43.09 |
| LUT | 164086 | %37.87 |
| Flip Flops | 50448 | %5.82 |

FPGA simulation was performed on Xilinx Virtex 7. One bottleneck of FPGA simulation is the generation of Additive White Gaussian Noise (AWGN) and feeding it into the LDPC decoder. This can be done using any high-level programming language. However, since the decoder is faster than both the noise generator program and the IO interface for transferring the encoded data, the slow speed of the encoder would add a significant delay to the input of the decoder. This will become the design's bottleneck when processing millions of codewords to measure BER and FER in higher SNR conditions. Likewise, the limited amount of available memory in the FPGA would not be sufficient for storing the pre-generated encoded data. In order to overcome this issue, an AWGN generator module inside the FPGA was deployed [18]. Doing so, with a clock frequency of 80 MHz and codewords with 864 LLRs, a throughput with the minimum of 2.86 Gbps (decoding the shorter code) and maximum of 3.06 Gbps (decoding the longer code) was obtained. It is worth noting that although decoding the shorter code results in a lower throughput compared to the longer code, more frames are processed at the similar time period. Clearly, having fewer parity bits leads to an increase in the number of iterations. However, for SNR bigger than 3dB the difference in the number of iterations compared to the longer code becomes negligible.

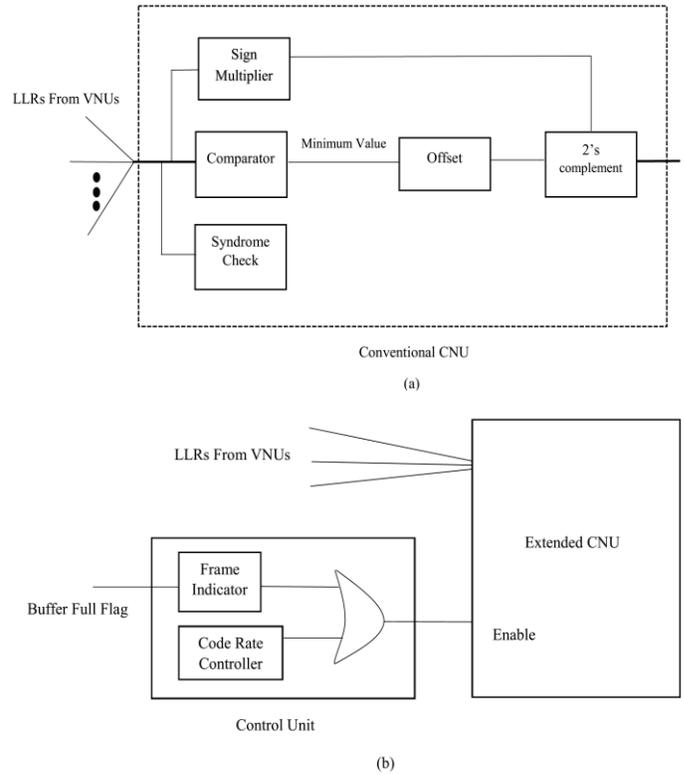

Fig. 6. Employed CNUs in the design. (a) Conventional CNU used for the mother code. (b) CNU with a control unit for extended parity.

V. CONCLUSION

A parity extension method with the goal of reducing the decoding latency in a fully parallel pipelined architecture is introduced. The extension is applied to a high rate WiMAX parity check matrix resulting in a lower rate parity matrix. The lower rate code is able to further increase the code performance in comparison to a standard WiMAX parity code with the same codeword size. In addition, the new parity code is able to close the performance gap with the 5G parity code. In our implementation, the decoder was capable of reducing decoding latency by 33% and consequently compensate for the initial delay associated with the pipelined architectures. Additionally, for SNR greater than 3 dB our design can reduce power consumption up to 36% by disabling unnecessary CNUs while processing the mother code.